\documentclass[sigconf]{acmart}
\setcopyright{none}
\renewcommand\footnotetextcopyrightpermission[1]{}

\AtBeginDocument{%
  }

\usepackage{amsmath,amsfonts,bm,pifont}
\usepackage[linesnumbered,ruled,vlined]{algorithm2e}
\usepackage{algpseudocode}
\usepackage{enumitem}

\newcommand{\model}{\texttt{DeSocial}\xspace}



\begin{document}

\title{From Aggregation to Selection: User-Validated Distributed Social Recommendation}




\author{Jingyuan Huang}
\affiliation{%
  \institution{Rutgers University}
  \city{Piscataway}
  \state{New Jersey}
  \country{USA}}
\email{chy.huang@rutgers.edu}

\author{Dan Luo}
\affiliation{%
  \institution{Lehigh University}
  \city{Bethlehem}
  \state{Pennsylvania}
  \country{USA}}
\email{danluo.ir@gmail.com}

\author{Zihe Ye}
\affiliation{%
  \institution{Rutgers University}
  \city{Piscataway}
  \state{New Jersey}
  \country{USA}}
\email{zihe.ye@rutgers.edu}

\author{Weixin Chen}
\affiliation{%
  \institution{Hong Kong Baptist University}
  \city{Hong Kong SAR}
  \country{China}
}
\affiliation{%
  \institution{Rutgers University}
  \city{Piscataway}
  \state{New Jersey}
  \country{USA}
}
\email{cswxchen@comp.hkbu.edu.hk}

\author{Minghao Guo}
\affiliation{%
  \institution{Rutgers University}
  \city{Piscataway}
  \state{New Jersey}
  \country{USA}}
\email{minghao.guo@rutgers.edu}

\author{Yongfeng Zhang}
\affiliation{%
  \institution{Rutgers University}
  \city{Piscataway}
  \state{New Jersey}
  \country{USA}}
\email{yongfeng.zhang@rutgers.edu}


\renewcommand{\shortauthors}{Jingyuan Huang et al.}


\begin{abstract}

Social recommender systems facilitate social connections by identifying potential friends for users. 
Each user maintains a local social network centered around themselves, resulting in a naturally distributed social structure.
Recent research on distributed modeling for social recommender systems has gained increasing attention, as it naturally aligns with the user-centric structure of user interactions.
Current distributed social recommender systems rely on automatically combining predictions from multiple models, often overlooking the user’s active role in validating whether suggested connections are appropriate.
Moreover, recommendation decisions are validated by individual users rather than derived from a single global ordering of candidates. As a result, standard ranking-based evaluation metrics make it difficult to evaluate whether a user-confirmed recommendation decision is actually correct.
To address these limitations, we propose \model, a distributed social recommendation framework with user-validation.
\model enables users to select recommendation algorithms to validate their potential connections, and the verification is processed through majority consensus among multiple independent user validators. To evaluate the distributed recommender system with user validator, we formulate this setting as a link prediction and verification task and introduce Acc@K, a consensus-based evaluation metric that measures whether user-approved recommendations are correct. Experiments on four real-world social networks demonstrate that \model improves decision correctness and robustness compared to single-point and distributed baselines. These findings highlight the potential of user-validated distributed recommender systems as a practical approach to social recommendation, with broader applicability to distributed and decentralized recommendations. Our code is available at: \url{https://github.com/agiresearch/DeSocial}.

\end{abstract}


\maketitle

\section{Introduction}

In the presence of massive volumes of information, recommender systems aim to select and rank items, such as friends, posts, or products, that are most relevant or valuable to users~\citep{yang2023dgrec,wang2022rete,chen2025dlcrec,TGOnline,GRU4Rec}. 
In social recommendation settings, such as friend suggestion, go beyond predicting user preferences; they also actively shape users’ personalized social circles~\citep{fan2019graph,burbach2019shares}. 
For instance, a user who initially connects with a small group of peers may be repeatedly exposed to recommendations involving their friends or frequent collaborators, gradually forming a dense local social neighborhood.
As this neighborhood evolves, users' observable behaviors, such as interaction patterns and subsequent connection choices, are strongly influenced. 
Therefore, modeling such evolving social neighborhood is essential for learning accurate and personalized user representations.

The reliance on local social neighborhoods motivates the design of distributed recommender systems, which preserve and leverage localized decision-making~\citep{chen2018privacy,wang2024would,krasanakis2022p2pgnn}.
Unlike conventional recommender systems that \textit{aggregate} user data and model outputs through a fixed pipeline~\citep{SLMRec,LightGCN,GRU4Rec,ying2021transformers,ROLAND}, distributed recommender systems offer an alternative paradigm in which multiple users or agents jointly influence recommendation outcomes~\citep{cai2024distributed,li2024federated,lu2025gradients,lin2020fedrec,zhang2024generative}. 
In particular, distributed recommender systems generate predictions by aggregating diverse judgments from multiple participating users, each equipped with different prediction models, and the target user makes decisions based on these collective judgments~\citep{user_control1,wang2024would}.
Thus, distributed recommender systems can provide better transparency, enhanced user controllability, and more trust~\citep{wang2024trustworthy,ge2024survey}.

However, existing distributed recommender systems is absent of \textbf{user validation}, causing them to degenerate into a \textbf{single-point, opaque, and weakly trusted aggregation mechanism}. Without user validation, recommendation decisions are made solely by automated aggregation functions. For example, the client models first produce item scores independently, which are then aggregated by a predefined rule ~\citep{he2021fedgraphnn,zhang2024e2gnn,chen2022ensemble,ammad2019federated,zheng2024poisoning} to generate the final recommendation list. This pipeline forms a purely machine-driven decision process, where the aggregation module becomes the single point of trust in an implicit way. 
Consequently, existing distributed recommender systems cannot explicitly incorporate user validation or trust into the decision process, relying instead on opaque aggregation mechanisms.
A further challenge lies in the \textbf{lack of evaluation metrics for recommender systems with user validation}. 
Standard metrics such as Hit@K, MRR, and NDCG are calculated based on a centralized setting, i.e., all evaluators share identical positive–negative samples and scoring functions~\citep{wang2022learning,LightGCN,schutze2008introduction,burges2005learning}. In contrast, user validation relies on multiple validators equipped with heterogeneous models and validation protocols, each providing only localized validation signals rather than item-level rankings. Consequently, those metrics is not directly applicable, which highlights the need for new evaluation measures that capture user validation and engagement.

To address the aforementioned challenges,  we propose framework \model, which specifically incorporates users validation into the recommendation decision process. 
Each user can select prediction model from a library of graph learning networks (e.g., MLP~\cite{MLP}, GCN~\cite{GCN}, GAT~\cite{GAT}, GraphSAGE~\cite{graphsage}, SGC~\cite{SGC}). During the selection, each user assesses candidate models using evaluation results derived from their local social neighborhood.
This user-centric validation process enables users to actively verify recommendation decisions based on their social circle, thereby introducing user-level trust signals into the aggregation process and enhancing transparency in distributed recommender systems.
After the validation, \model provides the final prediction through a majority consensus mechanism. 
To evaluate the distributed recommender systems with user validator, we propose a new metric named Acc@K. 
We formulate distributed recommendation as a graph link prediction task and we assess link prediction queries and determine final recommendations through a majority-based consensus mechanism.
In particular, Acc@K measures the probability that a user-approved positive link is ranked higher than $K\!-\!1$ sampled alternatives.
Therefore, Acc@K directly reflects whether the user-validated consensus leads to correct decisions, rather than merely higher-ranked items. 
Our contributions can be summarized as follows:
\begin{itemize}[leftmargin=*]
    \item We propose \model, a user-level distributed recommendation framework that incorporates users engagement into the decision-making loop. \model formulates recommendation as a graph link prediction task guided by user-validated multi-model consensus, enabling user control based on local social neighborhoods.

    \item We identify and formalize user-driven personalized algorithm selection as a key mechanism for enabling effective user validation in distributed social recommendation, and show that replacing aggregation with selection leads to more reliable decision outcomes under heterogeneous local views.

    \item We conduct extensive experiments on four real-world datasets, demonstrating that \model improves decision correctness and robustness compared to existing distributed and centralized baselines under user-involved recommendation settings.
\end{itemize}

\begin{figure*}[t]
\centering
\includegraphics[width=1\textwidth]{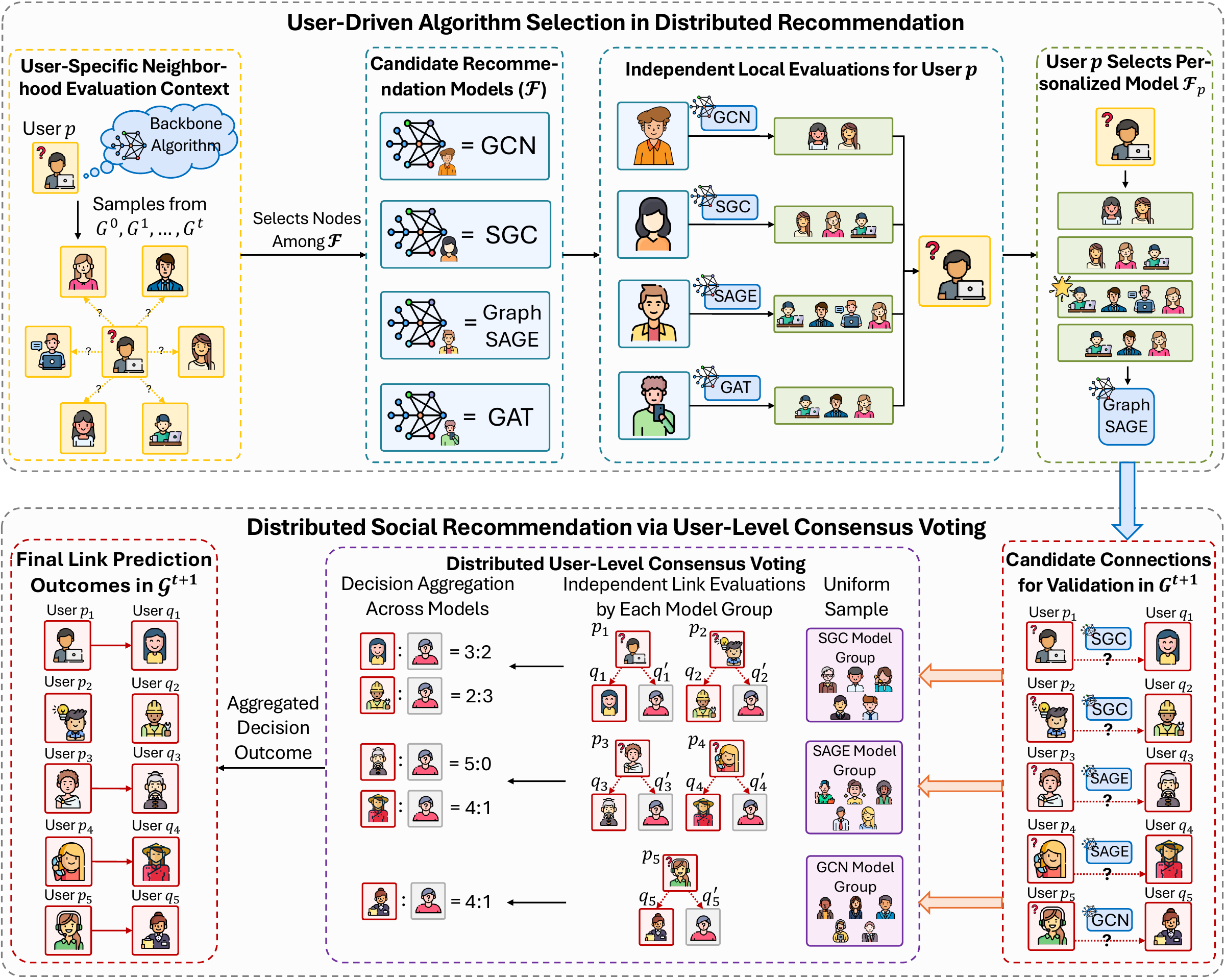}
\caption{
Illustration of \model, consisting of the user-driven algorithm selection and user-level consensus voting in distributed social recommendation.
Each user performs \emph{user-driven algorithm selection} by evaluating a pool of candidate recommendation models based on its local neighborhood, and selects a personalized model $\mathcal{F}_p$.
Given candidate connections in $\mathcal{G}^{t+1}$, validators independently conduct local inference using their selected models without shared parameters.
Final link predictions are obtained via \emph{distributed user-level consensus voting}, where binary judgments from validators are aggregated by majority voting.
}
\label{fig:PA}
\end{figure*}

\section{Preliminaries}

This work studies social recommendation where future connections at time $t+1$ are predicted based on historical social interactions, while accounting for user-specific preferences and local network structures at time $t$. 
In the following section, we introduce the notation used in this work and the formal definition of problems. 
%
%
We start from temporal graph:
\begin{definition}[\textbf{Temporal Graph}]
    A temporal graph can be formally introduced as $\mathcal{G}^{t}=(\mathcal{V}^t, \mathcal{E}^t)$ where $\mathcal{V}^t, \mathcal{E}^t$ denotes a set of $N$ users and a set of connections at time $t$, respectively.
\end{definition}


Since users may exhibit diverse behaviors and decision patterns, we further allow different users to employ different backbone models, which is referred to as user-specific backbones.

\begin{definition}[\textbf{User-Specific Backbone}]
    We denote the backbone used by user $u$, as $f_{\Theta _u}\in \mathcal{F}$ where $u$ is a user and $\Theta_u$ is the parameters of its backbone, $\mathcal{F}=\{\mathcal{F}_1, \mathcal{F}_2, ...\}$ is the backbone pool.
\end{definition}

In this work, each user-specific backbone serves as an validator that processes local graph information and produces a decision on observed connections. Based on the output of these validators, we define a vote as the basic decision signal used in our framework.

\begin{definition}[\textbf{Vote}]
    A vote is the boolean decision made by a validator $\phi$, typically another user, denoted as $\text{Vote}(\phi, p, q, t)\in\{0, 1\}$, indicating whether $\phi$ agree with the connection $(p,q)\in \mathcal{G}^t$.
\end{definition}

Given votes produced by individual validators, we introduce verification as a mechanism that aggregates multiple votes to validate connections in the social network.

\begin{definition}[\textbf{Verification}]
    A verification by a group of validators $\Phi_{p,t}$ for validating all the connections $(p,q)$ starts from $p$ at time $t$, can be formally defined as:
    \begin{equation}
    \label{eq:votesum}
    \text{Ver}(\Phi_{p,t}, p, q, t)=\text{Majority}(Vote(\phi, p, q, t)), \phi\in \Phi_{p,t},
    \end{equation} 
    \noindent 
    where $\text{Majority}(\cdot)$ is the majority of a list of decisions.
\end{definition}

Verification aggregates individual judgments into collective decisions on network connections. These collective decisions define the evolving graph structure observed by users. 
Now we are ready to formulate a distributed learning system in which users participating in verification jointly contribute to temporal prediction, while maintaining independent model parameters.

\begin{definition}[\textbf{User-Level Distributed Learning on Temporal Graphs}]
    User-level distributed learning involves all validator users at current time period collectively participating in the prediction of the graph structure for the next time step, is formulated as two steps:
\end{definition}

\begin{equation}
\{\Theta_{u}^{t-1}|u\in \mathcal{V}_{val}^t\} \xrightarrow{\text{test}}\mathcal{G}^{t},
\end{equation}

\begin{equation}
\min \left( \text{Loss}(\mathcal{\hat{G}}^{t},\mathcal{G}^{t};\Theta_{u}^t) \right), \ \forall u\in \mathcal{V}_{val}^t.
\end{equation}

Instead of optimizing a centralized model $\Theta^t$ and test the next period of graph $\mathcal{G}^{t}$, decentralization utilizes model parameters of all users to test $\mathcal{G}^{t}$, and all validator users $u\in \mathcal{V}_{val}^t$ in current period $t$ optimize their model parameters $\Theta^t_{u}$ independently. Formally, $\mathcal{V}_{val}^t$ is defined as
$\mathcal{V}_{val}^t = \bigcup_{(p,q)\in \mathcal{G}^{t}} \Phi_{p,t}.$


Under this formulation, the set of validator users in period $t$ is determined by their participation in connection verification. This allows model training and validation to be distributed across users without relying on a centralized model.

\section{\model Framework}
\label{section:proposal}

\subsection{Framework Overview}

As illustrated in \autoref{fig:PA}, \model consists of two tightly coupled components:
1) user-driven algorithm selection, which allows each user to choose the recommendation model that best fits their local social relationship, 
and 2) user-level consensus voting, where multiple peers validate each recommendation independently before it is finalized. This design is motivated by the following three challenges: 
\begin{itemize}[leftmargin=*]
    \item \textbf{No global optimization}: There is no centralized end-to-end training because of the distributed storage of social connctions and recommendation models.
    \item \textbf{No shared parameters}: Validators may use different recommendation algorithms with different inductive biases.
    \item \textbf{Consensus under variance}: Final recommendations must tolerate noise, diversity among validators.
\end{itemize}

In particular, a user $p$ who sends the connection validation request first selects a personalized recommendation model $\mathcal{F}_p$ from a candidate pool $\mathcal{F}$.
This selection is performed by evaluating candidate models on the historical neighborhood of user $p$, measuring how well each model distinguishes observed social connections from unobserved ones. 
By grounding model selection entirely on local neighborhoods, \textbf{\model avoids any form of global end-to-end optimization, making it compatible with distributed social graph storage}.

To generate predictions for the next snapshot $\mathcal{G}^{t+1}$, \model then distributes the request to a set of $n$ sampled validators. Each validator applies the same backbone model $\mathcal{F}_p$, but performs inference locally on its own subgraph, producing an independent recommendation judgment.
Importantly, \textbf{validators operate without shared parameters or synchronized training}, allowing recommendation models with different inductive bias to coexist in the system.

The final prediction is obtained by aggregating judgments from validators through majority voting. Rather than enforcing agreement at the model or parameter level, \textbf{\model reaches consensus at the user level, which enables robustness to noise and diversity across validators}. This design reflects a common recommendation principle: predictions are more reliable when they are supported by multiple independent but structurally related perspectives, rather than a single centralized model. In the following sections, we elaborate the user-driven algorithm selection and  user-level consensus voting, respectively. 


\subsection{User-Driven Algorithm Selection}

User-driven algorithm selection is designed to allow each user to independently select a personalized social recommendation model based on local neighborhood evaluation, without relying on global optimization.
Algorithm~\ref{alg:veri_algo} illustrates that each user is allowed to select a social recommendation algorithm independently.

\begin{algorithm}
    \renewcommand{\baselinestretch}{0.6}\normalsize
    \raggedright
    \SetAlgoLined
    \LinesNotNumbered
    \caption{User-Driven Algorithm Selection \label{alg:veri_algo}}
    \KwIn{Backbone pool set $\mathcal{F}$, neighborhood set size $\gamma$, adjust coefficient $\alpha$, and graph data $\mathcal{D}^{t}$}
    \KwOut{Personalized algorithm list $\mathcal{Q}$}

    \For{$u\in \{u|(u,v)\in \mathcal{G}^{t+1}\}$}{
        Calculate $\Gamma$ by Eq.~\ref{eq:neighborsample};

        \For{$(v_p,v_n)\in \Gamma$}{
            Calculate $\Pi_{u,v_p}$ given $t_e$ and $\alpha$;
        }

        $u$ calls users $r_1,r_2,...,r_{|\mathcal{F}|}$ that  contains different algorithms;

        \For{$i\in \{1,2,...,\mathcal{|F|}\}$}{
        
            \For{$(v_p,v_n)\in \Gamma$}{
    
                $r_i$ calculates $\mathbf{z_u, z_{v_p}, z_{v_n}}$ given $f_{\Theta_{r_i}}(\mathcal{D}^{t})$;
    
                $r_i$ calculates the probability of $(u,v_p)$ and $(u,v_n)$;

            }

            $r_i$ reports the weighted sum of the probability comparison to $u$;
            
        }

        $u$ selects $\mathcal{F}_u$ by Eq.~\ref{eq:selection_rule};
    }
    
    \Return $\mathcal{Q} \leftarrow \{\mathcal{F}_u|\exists v \in \mathcal{V},(u,v)\in \mathcal{G}^{t+1}\}$
\end{algorithm}

Given a set of recommendation algorithm $\mathcal{F}$, \model allows each node $u\in \mathcal{V}$ selects $\mathcal{F}_u\in \mathcal{F}$ for each $u$. Specifically, at time $t$, given a user $u$, we sample its positive and negative neighbor pairs \begin{equation}
\label{eq:neighborsample}
    \Gamma=\{(v_p, v_n)|(\bigcup_{\tau=0}^{t} v_p\in\mathcal{N}^t(u)) \bigwedge (\bigcup_{\tau=0}^{t} v_n \notin \mathcal{N}^t(u))\}
\end{equation} 
\noindent
with a size of $\gamma$, where positive samples correspond to users with historical interactions, and negative samples correspond to users with no historical interactions.

Afterwards, the user $u$ selects recommendation algorithm $\mathcal{F}_u$ through Eq.~\ref{eq:selection_rule}, where $\Pi_{u,v_p}$ denotes the edge weights for algorithm selection. The connection that emerges later has greater weight. $t_e$ is the emerge time of $(u,v)$, and $\alpha$ is the adjust coefficient. $\mathcal{P}(u,v)$ denotes the predicted probability that a connection exists between user $u$ and $v$ given the model $f$ trained by $\mathcal{D}^{t}$.

\begin{equation}
\label{eq:selection_rule}
\mathcal{F}_u = \arg\max_{f\in \mathcal{F}} \sum_{(v_p,v_n)\in\Gamma} 
\mathbb{I}\!\left[\mathcal{P}(u, v_p) > \mathcal{P}(u, v_n)
\right] \cdot \Pi_{u,v_p}
\end{equation}

\begin{equation}
\label{eq:edge_weights}
    \Pi_{u,v_p}=\exp(\alpha*(t-t_e))
\end{equation}

By leveraging the historical interactions, users can choose models that best fit their social relationship. In our implementation, the neighbor evaluation is performed using models that were already trained on the previous social connections. This design enables efficient and adaptive personalization while preserving computational efficiency. After the users select $\mathcal{F}_u$, \model starts to aggregate the decisions of different validator groups.

\subsection{User-Level Consensus Voting}

The voting and aggregation process is conducted by Algorithm~\ref{alg:pred_algo}.
\model predicts whether a connection $(u,v)$ will be formed in $\mathcal{G}^{t+1}$ through a user-level consensus voting mechanism among multiple validators.
It is particularly designed since the prediction of a single validator could be affected by local noise or model variance. 
With multiple validators, errors introduced by individual validators can be alleviated via majority voting.


\paragraph{Validation Committee Sampling}
To predict a social connection $(p_i,q_i)$ at time $t+1$, \model selects $n$ users using $\mathcal{F}_{p_i}\in \mathcal{F}$ and form a validation committee $\Phi_{p_i,t+1}$, as defined in Eq.~\ref{eq:committee}, where $\mathcal{V}_{\mathcal{F}_{p_i}}$ denotes the users using $\mathcal{F}_{p_i}$ as their recommendation algorithm. $\Phi_{p_i,t+1}$ is fixed at $t$ and specified by $\mathcal{F}_{p_i}$.

\begin{equation}
    \label{eq:committee}
    \Phi_{p_i,t+1}\sim \text{UniformSample}(\mathcal{V}_{\mathcal{F}_{p_i}},n)
\end{equation}

\paragraph{Local Inference}
To tackle the restriction of no global optimization, each selected validator $\phi_j\in \Phi_{p_i,t+1}$ copies $\mathcal{D}^t$ to its local memory to run $\mathcal{F}_{\phi_j}$ independently. Each $\phi_j$ predicts the connections $(p_i,q_i)$ and $(p_i,q_i'), q_i'\in Neg(\phi,p_i,q_i,t)$, making a binary decision $Vote(\phi_j,p_i,q_i,t+1)$ through Eq.~\ref{eq:vote}.  
At time $t$, each validator $\phi$ uses its recommendation model $f_{\Theta_\phi}$, as well as its current interactions $\mathcal{G}^t$, to compute model output vectors of the initiator $\mathbf{z_p^T}$ and the target $\mathbf{z_q}$. 
We also sample a set of users, $Neg(\phi, p, q, t)$, to indicate false connections $(p,q')$ where $q'\in Neg(\phi, p, q, t)$ for the comparison of prediction probability.
The decision can be interpreted as the comparisons among the cosine similarity of $\mathbf{z_p^T}$ and $\mathbf{z_r}, r\in \{q\}\cup Neg(\phi, p, q, t)$, which can be formally defined as
\begin{equation}
\label{eq:vote}
Vote(\phi,p,q,t) = \mathbb{I}\!\left[\mathcal{P}(p,q) > \max_{q'\in Neg(\phi,p,q,t)} \mathcal{P}(p,q') \right]
\end{equation}
\noindent
where $\mathcal{P}(p,q)$ denotes the predicted probability that a connection exists between $u$ and $v$, given $f_{\Theta_\phi}$ and historical interaction $\mathcal{D}^t=\bigcup_{\tau=0}^{t}\mathcal{G}^{\tau}$. When a user $u$ request a connection validation on $(u,v)$ at time $t$, \model selects a set of validators $\Phi$. Each validator $\phi_i \in \Phi$ makes a decision $Vote(\phi_i, u, v, t)$ based on its recommendation model $f_{\Theta _{\phi_i}}=f_{\Theta _u}$ and historical interactions $\mathcal{D}^t$.

\begin{algorithm}
    \renewcommand{\baselinestretch}{0.1}\normalsize
    \raggedright
    \SetAlgoLined
    \LinesNotNumbered
    \caption{User-Level Consensus Voting\label{alg:pred_algo}}
    \KwIn{Social graph data $\mathcal{D}^{t}$, validator set size $n$, randomly initialized parameters of each user $\Theta_u$, recommendation algorithm pool $\mathcal{F}$, request batch size $B_{\text{req}}$, voting batch size $B_{\text{vote}}$.}
    \KwOut{Model parameters $\{\Theta_{u}^{\star}|\ u\in \mathcal{V}\}$}

    \For{time period $t \in \{0,1,...,T-1\}$}{

        \textbf{\# Validation committee sampling}

        \For{$p\in \{p|\exists q\in \mathcal{V} , (p,q)\in \mathcal{G}^{t+1}\}$}{
            Compute $\Phi_{p,t+1}$ via Eq.~\ref{eq:committee};
            
            $\mathcal{Q}_p^{t+1}\leftarrow \{q|(p,q)\in \mathcal{G}^{t+1}\}$;

            \For{each batch $\mathcal{B}\in \text{Partition}(\mathcal{Q}_p^{t+1}, B_{\text{req}})$}{
                $p$ request validation of $\mathcal{B}$ from users in $\Phi_{p,t+1}$;
            }
        }

        Calculate $\mathcal{V}_{val}^t$ by taking the union of all $\Phi_{p,t+1}$;

        \textbf{\# Local inference}

        \While{$\exists\ u\in \mathcal{V}_{val}^t$ such that $\Theta_{u}$ does not converge}{
            
            \For{$u\in \mathcal{V}_{val}^t$}{
                Predict the future graph $\mathcal{\hat{G}}^{t+1}$ given $\Theta_{u}$ and $\mathcal{D}^t$;
                
                Calculate the prediction loss $\mathcal{L}(\mathcal{\hat{G}}^{t+1}, \mathcal{G}^{t+1})$;
                
                Optimize $\Theta_{u}$ using $\mathcal{L}(\mathcal{\hat{G}}^{t+1}, \mathcal{G}^{t+1})$;
            }
        }

        \textbf{\# Aggregation and voting}

        \For{each batch $\mathcal{B'}\in \text{Partition}(\mathcal{G}^{t+1}, B_{\text{vote}})$}{
    
            \For{$(p,q)\in \mathcal{B'}$}{
            
                \For{$u\in \Phi_{p,t+1}$}{
                    $u$ gives $Vote(\Phi_{p,t+1}, p, q, t+1)$;
                }
        
                Aggregate votes and compute 
                $Ver(\Phi_{p,t+1}, p, q, t+1)$ via Eq.~\ref{eq:votesum};
            }
        }
    }

    $\forall\ u\in \mathcal{V}, \Theta_{u}^{\star} \leftarrow \Theta_{u}$;
    
    \Return $\{\Theta_{u}^{\star}|\ u\in \mathcal{V}\}$
\end{algorithm}

\paragraph{Aggregation} 
The system coordinates a vote collection process among the selected committee members $\phi_1, \phi_2, \dots, \phi_n$.
Each validator independently returns a binary judgment $Vote(\phi_j, p_i, q_i, t+1)$ based on its local inference.
The final decision is obtained by aggregating these votes through majority rule.
As shown in Eq.~\ref{eq:votesum}, a positive recommendation is produced if more than half of the committee agrees, i.e.,
$\sum_{i=1}^n Vote(\phi_j, p_i, q_i, t+1) > \lfloor \frac{n}{2} \rfloor$.

\paragraph{Batched Request and Voting}
We partition connection validation requests into batches of size $B_{\text{req}}$ to improve the efficiency of distributed inference and communication. Similarly, voting is conducted over connection batches of size $B_{\text{vote}}$, reducing per-connection evaluation overhead.

\section{Experimental Setups}

\subsection{Experimental Protocol}
Following common practice in sequential prediction tasks, we adopt a rolling leave-one-out evaluation protocol.
Here, leave-one-out refers to leaving out the next temporal slice for evaluation, rather than removing individual users or connections.
At time period $t$, each validator trains its model independently using historical interactions
$\bigcup_{\tau=0}^{t-1}\mathcal{G}^{\tau}$, validation is conducted on $\mathcal{G}^t$, and testing on $\mathcal{G}^{t+1}$.
This design provides a consistent and controlled training protocol across validators, allowing us to isolate the effects of distributed validation and consensus aggregation from those of incremental or adaptive optimization strategies.
We restrict model updates to users selected as validators for $\mathcal{G}^{t+1}$.
This choice aligns with the execution pattern of \model, where only participating validators actively engage in recommendation validation at each time step.
While \model could in principle update all nodes using $\mathcal{G}^t$, such global updates are unnecessary for evaluating consensus behavior and would introduce substantial computational overhead in a centralized simulation.
By limiting updates to the active validator set, we retain the essential dynamics of distributed validation while ensuring scalable and reproducible experimentation. 
Our experiments are conducted on a server equipped with an Intel Xeon Gold 6226R CPU and eight NVIDIA A100 GPU. 

\subsection{Datasets}

We applied four real-world datasets (\textbf{UCI}~\cite{poursafaei2022towards}, \textbf{Memo-Tx}~\cite{zuo2023set}, \textbf{Enron}~\cite{zhang2024dtgb}, and \textbf{GDELT}~\cite{zhang2024dtgb}). For each dataset, we divide the social interactions into a sequence of 40 discrete time slices by uniformly partitioning all edges according to their timestamps, ensuring that each time slice contains approximately the same number of interactions while preserving the overall temporal order of events. Detailed information on the datasets is described in Appendix \ref{sec:dataset}.

\begin{table*}[t]
\centering
\caption{Performance comparison of \model and baselines on four datasets in terms of Acc@2 and Acc@3 (\%). The mean and standard deviation across five validator settings are reported. The highest scores are in \textbf{bold}, and the second-highest are underlined. Acc@5 results are reported in \autoref{fig:acc5}.}
\label{tab:main_res}
\setlength{\tabcolsep}{6pt}
\begin{tabular}{c|cc|cc|cc|cc}
\hline
\textbf{Dataset} & \multicolumn{2}{c|}{\textbf{UCI}} & \multicolumn{2}{c|}{\textbf{Memo-Tx}} & \multicolumn{2}{c|}{\textbf{Enron}} & \multicolumn{2}{c}{\textbf{GDELT}} \\
\hline
\textbf{Metrics} & \textbf{Acc@2} & \textbf{Acc@3} & \textbf{Acc@2} & \textbf{Acc@3} & \textbf{Acc@2} & \textbf{Acc@3} & \textbf{Acc@2} & \textbf{Acc@3} \\
\hline
\textbf{MLP}        & 66.38$\pm$0.34 & 52.52$\pm$0.30 & 73.61$\pm$0.11 & 66.57$\pm$0.11 & 81.48$\pm$0.08 & 75.20$\pm$0.09 & 91.28$\pm$0.02 & 87.60$\pm$0.02 \\
\textbf{GCN}        & 63.90$\pm$0.17 & 51.90$\pm$0.19 & 69.62$\pm$0.13 & 61.23$\pm$0.76 & 79.92$\pm$0.09 & 74.41$\pm$0.09 & 82.94$\pm$0.04 & 74.08$\pm$0.06 \\
\textbf{GAT}        & 61.15$\pm$0.26 & 48.24$\pm$0.28 & 72.51$\pm$0.26 & 65.88$\pm$1.07 & 85.52$\pm$0.12 & 80.30$\pm$0.14 & 88.29$\pm$0.28 & 81.34$\pm$0.39 \\
\textbf{GraphSAGE}       & 69.00$\pm$0.40 & 55.78$\pm$0.48 & 82.85$\pm$0.15 & 75.47$\pm$0.16 & \underline{90.27$\pm$0.06} & \underline{86.16$\pm$0.07} & 93.16$\pm$0.02 & 89.59$\pm$0.03 \\
\textbf{SGC}        & 72.77$\pm$0.24 & 62.77$\pm$0.24 & 80.37$\pm$0.05 & 74.78$\pm$0.07 & 88.24$\pm$0.04 & 84.50$\pm$0.08 & \underline{95.59$\pm$0.02} & \underline{92.46$\pm$0.02} \\
\hline
\textbf{GEN}        & 66.19$\pm$2.20 & 52.42$\pm$2.91 & 81.92$\pm$1.21 & 74.41$\pm$1.38 & 90.04$\pm$0.65 & 85.77$\pm$0.77 & 92.87$\pm$0.17 & 83.94$\pm$0.26 \\
\textbf{E2GNN}      & \underline{74.98$\pm$0.69} & \underline{63.00$\pm$1.17} & \underline{83.83$\pm$0.65} & \underline{76.83$\pm$0.53} & 86.44$\pm$0.17 & 80.65$\pm$0.17 & 94.95$\pm$0.06 & 91.34$\pm$0.07 \\
\textbf{FedGraphNN} & 70.47$\pm$0.51 & 60.08$\pm$0.43 & 82.72$\pm$0.28 & 75.32$\pm$0.28 & 88.87$\pm$0.13 & 84.39$\pm$0.10 & 95.50$\pm$0.03 & 92.33$\pm$0.04 \\
\hline
\textbf{\model}    & \textbf{77.63$\pm$0.36} & \textbf{66.01$\pm$0.44} & \textbf{87.25$\pm$0.17} & \textbf{79.65$\pm$0.16} & \textbf{92.11$\pm$0.13} & \textbf{88.39$\pm$0.13} & \textbf{98.13$\pm$0.03} & \textbf{96.09$\pm$0.04} \\
\hline
\end{tabular}
\end{table*}

\begin{table}[t]
\caption{Ablation studies of \model on UCI, Memo-Tx, and Enron in terms of Acc@2 and Acc@3 (\%). The best results are in bold, and the second-best are underlined.}
\label{tab:ablation}
\small
\centering
\setlength{\tabcolsep}{2pt}
\begin{tabular}{c|cc|cc|cc}
\hline
\textbf{Datasets} & \multicolumn{2}{c|}{\textbf{UCI}} & \multicolumn{2}{c|}{\textbf{Memo-Tx}} & \multicolumn{2}{c}{\textbf{Enron}} \\
\hline
\textbf{Metrics} & \textbf{Acc@2} & \textbf{Acc@3} & \textbf{Acc@2} & \textbf{Acc@3} & \textbf{Acc@2} & \textbf{Acc@3} \\
\hline
\textbf{Single Backbone} & 72.77 & 62.77 & 82.85 & 75.47 & 90.27 & 86.16 \\
\hline
\textbf{\model} & \textbf{77.63} & \textbf{66.01} & \textbf{87.25} & \textbf{79.65} & \underline{92.11} & \textbf{88.39} \\ \hline
\textbf{w/o Personalized} & \underline{76.28} & \underline{65.64} & 85.88 & \underline{79.20} & \textbf{92.18} & \underline{88.19} \\
\textbf{Random Select} & 75.05 & 61.10 & 84.81 & 78.40 & 90.83 & 87.11 \\
\textbf{Simple Select} & 73.33 & 56.66 & \underline{86.44} & 78.13 & 90.67 & 87.06 \\ \hline
\textbf{w/o Consensus} & 73.35 & 63.16 & 83.96 & 77.28 & 90.08 & 86.36 \\ \hline
\end{tabular}
\end{table}

\subsection{Baselines}

We used five classic graph models (\textbf{MLP}~\cite{MLP}, \textbf{GCN}~\cite{GCN}, \textbf{GAT}~\cite{GAT}, \textbf{GraphSAGE}~\cite{graphsage}, and \textbf{SGC}~\cite{SGC}) in social recommendation. We also compared with other open-sourced distributed frameworks, such as ensemble learning (\textbf{GEN}\cite{duan2024graph} and \textbf{E2GNN}~\cite{zhang2024e2gnn}) and federated learning  (\textbf{FedGraphNN}\cite{he2021fedgraphnn}), which supports these five models as their backbones. We report a detailed description for these baselines and their setups in Appendix ~\ref{sec:baseline} and Appendix ~\ref{sec:imp_detail}. Appendix ~\ref{sec:imp_detail} also shows implementation details of \model.



\subsection{Evaluation Metrics}

Conventional recommendation metrics, such as Hit@K, MRR, and NDCG, assume a single scoring function producing a global item ranking.
In contrast, our framework relies on a majority voting consensus among multiple user validators, each providing heterogeneous models and validation protocols, rather than full rankings.
As a result, these metrics are not directly applicable in our setting.
To address this issue, we formulate distributed recommendation as a graph link prediction task and adopt \textbf{Acc@K} as the evaluation metric.
Specifically, given one user-approved positive connection and $K-1$ sampled negative connections, \textbf{Acc@K} measures the probability that the positive connection receives the highest consensus score among the $K$ candidates.
We evaluate with $K\in\{2,3,5\}$, where larger $K$ corresponds to more challenging decision scenarios.
Negative connections are sampled using the randomized strategy proposed in~\cite{zhang2024dtgb}.

\section{Experiments}

We conduct extensive experiments to demonstrate the effectiveness of our \model by investigating the following research questions:


\begin{itemize}[leftmargin=*]
    \item \textbf{RQ1}: How does the performance of \model compare with existing distributed and single-point recommender systems?
    \item \textbf{RQ2}: What impact do different design variants have on \model, particularly regarding the effectiveness of personalized algorithm selection and user-driven algorithm selection?
    \item \textbf{RQ3}: How can user validation and full agreement among validators be properly measured in distributed social recommendation?
    \item \textbf{RQ4}: How does validator selection influence the effectiveness of user validation in distributed social recommendation?
\end{itemize}

\subsection{Main Results (RQ1)}

The overall performance comparison of \model with single-backbone models, ensemble methods, and federated learning baselines is reported in \autoref{tab:main_res} and \autoref{fig:acc5}.
Across all datasets, \model outperforms the strongest single-backbone baseline, achieving an average improvement of 4.51\%.
More importantly, \model exhibits clear advantages over aggregation-based approaches.
Compared with ensemble methods such as GEN and E2GNN, which rely on parameter fusion or soft voting, \model adopts a fundamentally different decision mechanism.
Instead of aggregating scores or model parameters, \model performs decision-level selection through user-validated multi-model consensus.
The superior performance of \model indicates that social graph link prediction cannot be effectively solved by aggregation alone, even when multiple strong models are available.
Furthermore, the comparison between \model and aggregation-based federated methods highlights the limitations of distributed aggregation.
Although FedGraphNN enables distributed training, its clients operate on partial graph views and contribute model updates through predefined aggregation rules.
In contrast, \model empowers validators with global graph awareness to assess candidate connections independently and participate in the final decision via consensus.
This shift from model aggregation to user-validated selection leads to more reliable recommendation outcomes.
Overall, these results show that \textbf{user-driven personalized algorithm selection is essential for effective validation in distributed social recommendation, as aggregation alone fails to capture heterogeneous local evidence across users}.






\subsection{Ablation Studies (RQ2)}

\begin{figure*}[ht]
\centering
\includegraphics[width=1\textwidth]{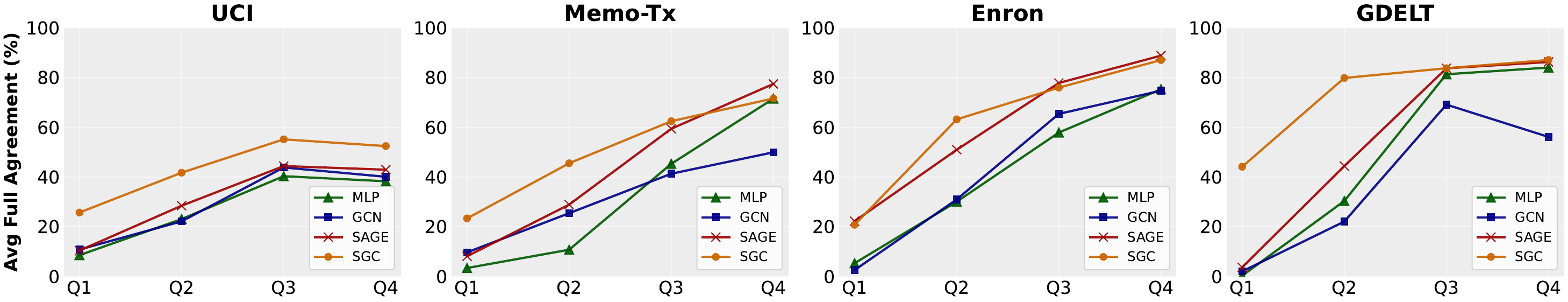}
\caption{Full agreement across degree node quartiles. By incorporating user validation, least active (Q1) users can achieve improved outcomes, as decisions are informed by the collective assessments of active validators rather than sparse local data.}
\label{fig:quartile}
\end{figure*}


In this section, we investigate the impact of personalized algorithm
selection and user-driven algorithm selection as we proposed in Section~\ref{section:proposal}.
Our experimental results are summarized in \autoref{tab:ablation}.
In particular, we conduct the following ablation studies to examine the contribution of each module in \model:  

\begin{itemize}[leftmargin=*]
    \item \textbf{w/o Personalized}: Only a fixed algorithm is considered, with predictions made by five validators using the same model.  
    \item \textbf{Random Select}: Each node selects an algorithm from $\mathcal{F}$ uniformly at random.  
    \item \textbf{Simple Select} (see Appendix~\ref{sec:rule-based-sel}): Algorithm choice is determined by two local structural features, namely node degree and clustering coefficient.  
    \item \textbf{w/o Consensus}: Each validator community reduces to a single validator making the final decision without consensus voting.  
\end{itemize}

\subsubsection{Personalized Algorithm Selection}

Contrasting with random and simple selection methods, \model allows each user to choose the most appropriate model for their local conditions consistently leads to superior performance.
This is because personalized algorithm selection enables reliable validation without shared parameters by allowing users to independently incorporate algorithms that align with their own structural features derived from local social neighborhoods. 
These results demonstrate that \textbf{user validation grounded in local social neighborhoods plays a critical role in improving decision correctness while avoiding parameter sharing.}

\subsubsection{Multi-User Consensus}

\begin{figure}[!t]
\centering
\includegraphics[width=0.36\textwidth]{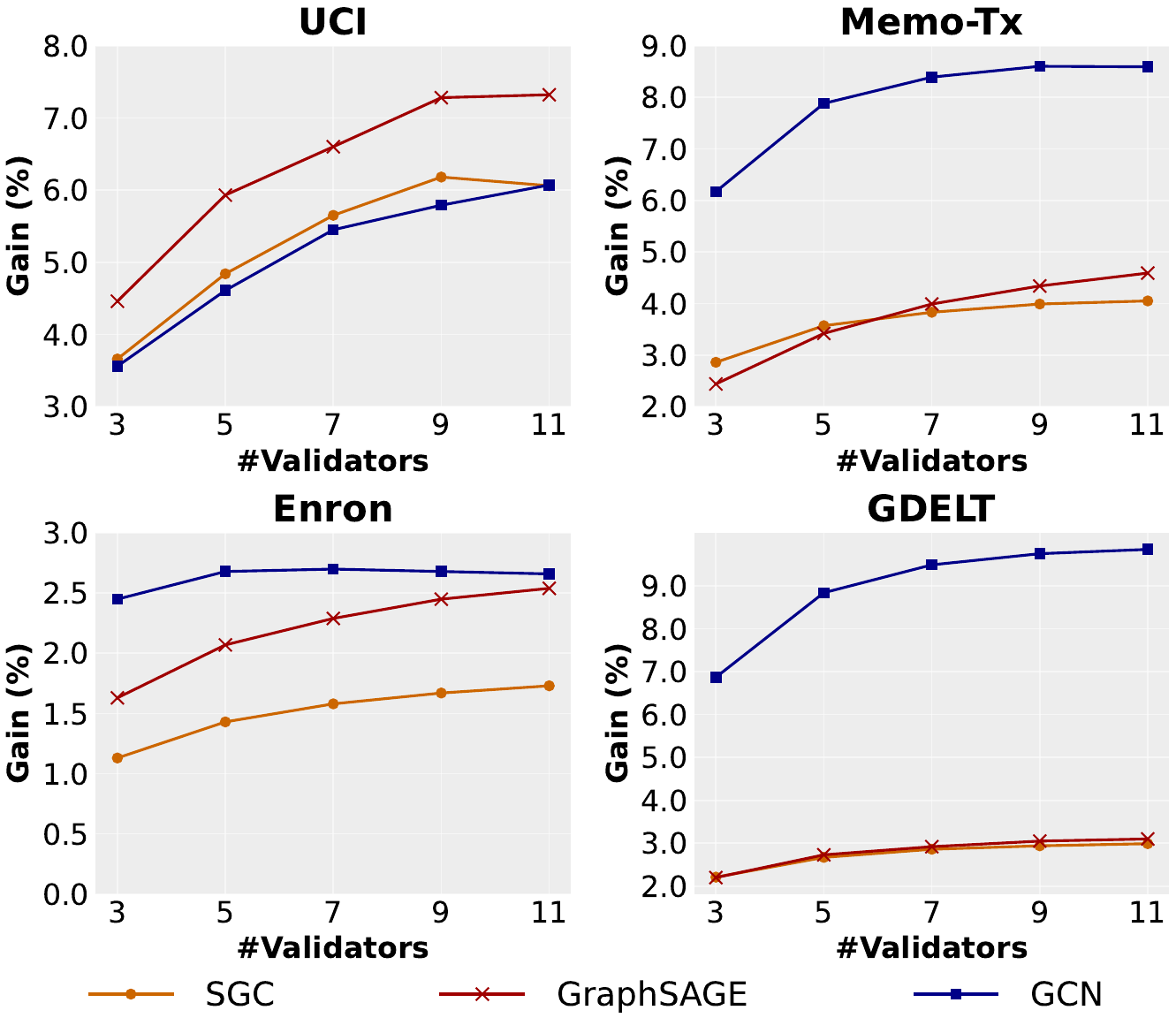}
\caption{Gains versus number of validators. We vary the validator committee size 
$n$ and report the corresponding average gain of Acc@2/3/5. Gains converges as $n$ increases.}
\label{fig:module2acrossdiffn}
\end{figure} 

Removing the consensus mechanism results in a significant performance drop, validating the effectiveness of majority voting among validators.
As long as the single-point algorithm outperforms random guessing, consensus provides considerable robustness against individual model variance.
The only exceptions occur, as shown in \autoref{fig:acc5_cons}, when the selected backbone exhibits performance below random, in which case aggregation tends to amplify errors.
This observation further demonstrates how user-driven algorithm selection induces effective consensus among heterogeneous validators \textbf{by ensuring that only sufficiently strong algorithms participate in validation, as weaker algorithms may otherwise lead to diminished robustness}.
We further analyze how the performance gain from multi-node consensus over a single recommendation algorithm varies with the number of validators as shown in \autoref{fig:module2acrossdiffn}.
As the committee size increases, the marginal improvement gradually diminishes, suggesting that practical committee sizes need not be very large.

\subsection{Full Agreement across User Activity Levels (RQ3)}

To examine how user validation can be properly measured under different activity levels, we divide users into four quartiles based on the number of connections, from the least (Q1) to the most (Q4). We regard the users with more connections as more active users.
\autoref{fig:quartile} reports \textbf{the proportion of full agreement at Acc@2} for each quartile (i.e., of all the agreements in this quartile, how many of them have all the validators given true votes), which we use as an explicit signal of user validation quality. Q1 exhibits the lowest proportion due to limited information of connections. Q2 and Q3 benefit from richer information, leading to more reliable predictions and higher proportion. In most cases, Q4 achieves the highest. However, on UCI and GCN on GDELT, Q4 shows slightly lower proportion than Q3, likely because redundant neighborhood information introduces conflicting structural signals, increasing disagreement among validators.
Overall, consensus signals are primarily contributed by more active users with richer neighborhoods. These reliable validation signals can then support predictions for sparsely connected nodes. Therefore, \textbf{full agreement provides an interpretable signal of user validation quality, revealing how consensus reliability varies across activity levels and why standard ranking-based metrics fail to capture this effect}.

\begin{figure*}[ht]
\centering
\includegraphics[width=1\textwidth]{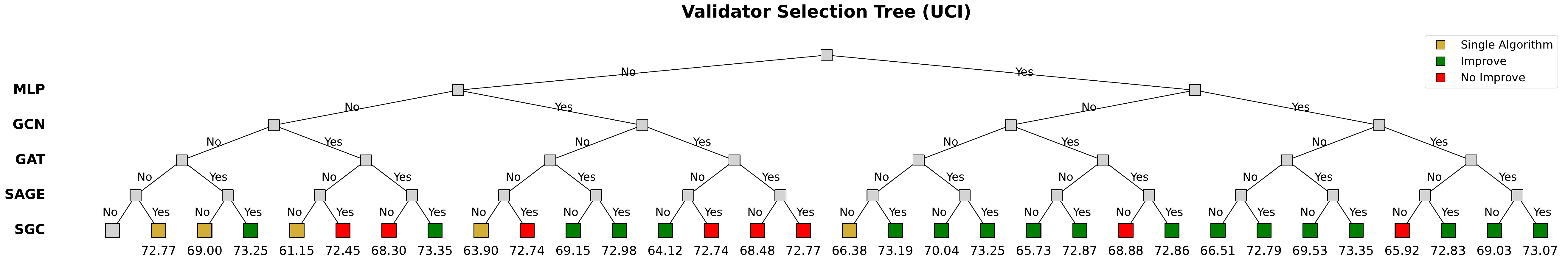}
\caption{Validator selection trees illustrating how different types of validator selection affect user validation on UCI. Some of the recommendation algorithm combinations can improve the performance (Acc@2) while others degrades.}
\label{fig:pa_uci}
\end{figure*}

\begin{figure*}[ht]
\centering
\includegraphics[width=1\textwidth]{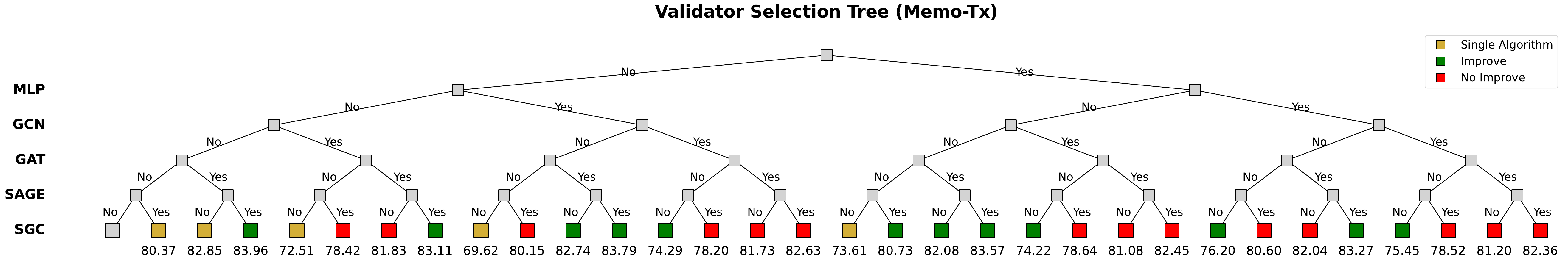}
\caption{Validator selection trees illustrating how different types of validator selection affect user validation on Memo-Tx. The valid type of validator selection (with Acc@2 improved) varies, depending on the datasets.}
\label{fig:pa_memo}
\end{figure*}

\subsection{Selecting the Right Validators (RQ4)}

To further understand how validator selection influences the effectiveness of user validation, we visualize the validation outcomes shown in \autoref{fig:pa_uci} and \autoref{fig:pa_memo}. Due to space limitations, we visualize the results with Acc@2, while similar trends are observed for other Acc@K.
Each root-to-leaf path represents a type of validator selection, combining different recommendation algorithms, and leaf nodes indicate Acc@2.
Green and red leaves denote improved and degraded performances, respectively, highlighting that validator selection plays a critical role in distributed social recommendation.
The trees reveal a ``less-is-more'' phenomenon: incorporating additional recommendation models may not  improve the validation and may even harm it.
This suggests that effective user validation relies on selecting appropriate validators with complementary inductive biases, rather than blindly increasing different kinds of participating models.
In summary, \textbf{these results demonstrate that validator selection is a key factor in enabling reliable user validation}.

\section{Related Works}

\subsection{Graph Recommender Systems}

Existing graph recommendation methods, including graph neural networks (GNNs)~\cite{GCN, GAT, graphsage, SGC, GIN, LightGCN, JODIE, wang2021inductive}, meta-learning~\cite{SML, IGC, ROLAND, MetaDyGcn, TGOnline}, Transformer-based~\cite{SGFormer, ying2021transformers}, and LLM-based approaches~\cite{GraphPrompter, LLaGA, he2023harnessing, yu2025graph2text, PromptGFM}, are primarily designed for single-point settings, where recommendation quality is driven by global representation learning and model optimization.
In contrast, we study distributed social recommendation, where users only observe local social neighborhoods and cannot share raw data~\cite{zhang2021federated, qi2022fairvfl, meng2021cross, qin2024blockdfl}.
Under this setting, reliable recommendation depends on user-level algorithm selection and validation rather than learning a single-point model.
In our study, we focus on lightweight graph models, including MLP~\cite{MLP}, GCN~\cite{GCN}, GAT~\cite{GAT}, GraphSAGE~\cite{graphsage}, and SGC~\cite{SGC}, which provide diverse inductive biases while remaining practical for distributed deployment.

\vspace{-0.3cm}
\subsection{Model Aggregation Methods}

Numerous aggregation methods have been proposed to combine predictions or information from multiple models.
Ensemble learning aggregates independently trained models to reduce overfitting or noise~\cite{dong2020survey, breiman1996bagging, freund1997decision},
and has been adopted in graph recommendation to mitigate sensitivity to graph topology by fusing multiple models~\cite{duan2024graph, zhang2024e2gnn, chen2022ensemble}.
Most aggregation methods rely on voting mechanisms to form a single prediction,
including soft voting that averages probabilistic outputs and hard voting that applies majority rules~\cite{wen2024black, ye2025pilot, franklin2011crowddb}.
Federated learning extends aggregation to distributed settings by combining model parameters or gradients without sharing raw data~\cite{mcmahan2017communication, he2021fedgraphnn, zhang2023secure}.
Similarly, peer-to-peer graph learning frameworks remove the central server to achieve decentralization, but still rely on parameter exchange among participants~\cite{krasanakis2022p2pgnn}.
While effective for collaborative training, these methods assume that aggregation over models or parameters is sufficient to produce reliable decisions.
In contrast, our work focuses on \emph{selection} rather than aggregation, enabling users to validate and choose recommendation algorithms based on local social neighborhoods when model reliability varies across users.

\section{Conclusions}


We study social recommendation in user-level distributed settings and propose \model, a framework that explicitly incorporates user validation into the recommendation decision process.
Instead of relying on centralized model aggregation, \model enables users to select personalized recommendation algorithms based on local social neighborhoods and determines final predictions through validator-level consensus.
This design improves robustness and flexibility in distributed environments by allowing heterogeneous models to coexist and be evaluated independently.
Through extensive experiments, we demonstrate that personalized algorithm selection plays a critical role in improving decision correctness, and that effective user validation depends on selecting appropriate validators rather than indiscriminately increasing their number.
Our analysis further reveals a clear ``less-is-more'' phenomenon in validator composition, highlighting the importance of model complementarity and selection.
Overall, \model provides a user-centric perspective on distributed social recommendation, offering a principled way to integrate personalized model selection and consensus-based validation.

\section{Limitations and Future Works}

Our study is limited to a fixed set of lightweight backbone models, which is sufficient to reveal the effects of personalized selection and validator composition, but does not capture more diverse or evolving model pools.
In addition, we do not explicitly model strategic or unreliable validator behavior, and assume validators make independent decisions based on local social neighborhoods.
Exploring adaptive validator selection under dynamic social structures and heterogeneous validator reliability remains an important direction for future works.

\bibliographystyle{ACM-Reference-Format}
\bibliography{references}

\appendix
\section{Appendix}

\subsection{Datasets}
\label{sec:dataset}

\autoref{tab:dataset} shows the number of nodes and edges, network density of each dataset.

\begin{table}[ht]
\centering
\renewcommand{\arraystretch}{0.9}
\caption{Dataset Statistics}
\label{tab:dataset}
\begin{tabular}{c|ccc}
\hline
\textbf{Dataset} & \textbf{\#Users} & \textbf{\#Interactions} & \textbf{Density}  \\ \hline
\textbf{UCI} & 1,899 & 59,835 & 0.016592 \\
\textbf{Enron} & 42,711 & 797,907 & 0.000437  \\
\textbf{GDELT} & 6,786 & 1,339,245 & 0.029083  \\
\textbf{Memo-Tx} & 10,907 & 994,131 & 0.008357  \\ \hline
\end{tabular}
\end{table}

\begin{itemize}[leftmargin=*]
    \item \textbf{UCI}~\cite{poursafaei2022towards} is a spatio-temporal network map of student interactions at UC Irvine, showing timestamped communication relationships between students.

    \item \textbf{Enron}~\cite{zhang2024dtgb} is a temporal graph dataset constructed from email communications between Enron employees from 1999 to 2002. The edges are ordered by sending time.

    \item \textbf{GDELT}~\cite{zhang2024dtgb} is a media-based temporal graph of global political and social events. Each node represents an entity (e.g., a person, country, or organization). Edges were arranged in chronological order, indicating that the entities were mentioned together.

    \item \textbf{Memo-Tx}~\cite{zuo2023set} is a Web 3.0 transaction graph from \texttt{memo.cash}, a blockchain-based social platform. We use timestamped transactions from 4/6/2018 to 11/30/2021, group them into blocks with multiple inputs/outputs. For each block, we connect every input to every output of the same cryptocurrency to represent currency flows.
\end{itemize}

\subsection{Baselines}
\label{sec:baseline}

We attach all the baselines as follows.

\begin{itemize}[leftmargin=*]

\item \textbf{MLP}~\cite{MLP} only uses node attributes, ignoring the topology of the graph. It is a powerful feature-based basis that is computationally efficient as it does not rely on the structure.

\item \textbf{GCN}~\cite{GCN} performs graph spectral convolution by aggregating information about the immediate neighbors' features. It efficiently captures local homophilic patterns.

\item \textbf{GAT}~\cite{GAT} introduces attention mechanism to assign adaptive weights to neighbors during aggregation. This is particularly effective on heterogeneous or noisy connected graphs, but has a high computational cost due to the attention computation.

\item \textbf{GraphSAGE}~\cite{graphsage} learns aggregation functions in sampled neighbors, enabling generalization to unseen nodes, suitable for large and dynamic graphs, balancing performance and scalability. It computes faster by reducing the aggregation neighbors.

\item \textbf{SGC}~\cite{SGC} removes nonlinearity and collapses multiple layers of graph convolution network into a single linear transformation with pre-computed propagation. It improves speed significantly while maintaining high performances on homogeneous graphs.

\item \textbf{GEN}~\cite{duan2024graph} averages node embeddings from different backbones to ensemble them, as they are complementary to each other and provide enhanced representations.

\item \textbf{E2GNN}~\cite{zhang2024e2gnn} alleviates the limitations of naive GNN ensembles by aggregating the predictions of multiple backbones with a lightweight MLP, and leveraging a reinforced discriminator to filter out noisy pseudo-labels, improving the accuracy and robustness.

\item \textbf{FedGraphNN}~\cite{he2021fedgraphnn} is a federated learning framework for GNNs where the data is partitioned and distributed over multiple clients that train models on their local subgraphs and aggregate the parameters together, for both better performance and data privacy.

\end{itemize}

\subsection{Simple Rule-Based Selection}
\label{sec:rule-based-sel}
To demonstrate the effectiveness of \model's personalized algorithm selection, we added an additional ablation study that each user simply selects a personalized backbone algorithm based solely on two trivial local subgraph features, specifically:
\begin{itemize}[leftmargin=*]
    \item Node Degree ($Deg(u)$): Defined as the number of 1-hop neighbors of user $u$ in the graph.
    \item Clustering Coefficient ($c_u$): Measures the density of the ego-network of user $u$, computed as the ratio between the number of connections among its neighbors and the maximum possible number of such connections.
\end{itemize}

Formally, given a graph $\mathcal{G}=(\mathcal{V}, \mathcal{E})$, let $\mathcal{N}(u)$ be the 1-hop neighbor set of node $u$, and $|\mathcal{N}(u)| = Deg(u)$. The clustering coefficient $c_u$ is calculated as:

\begin{equation}
\label{eq:clustering}
c_u =
\begin{cases}
0, & \text{if } Deg(u) \leq 1 \\
\frac{2 \cdot |\{ (v, w) \in \mathcal{E} \mid v, w \in \mathcal{N}(u) \}|}{Deg(u) \cdot (Deg(u) - 1)}, & \text{otherwise}
\end{cases}
\end{equation}

Algorithm \ref{alg:rule_selection} illustrates the simple backbone selection rules.

\begin{figure}[!t]
\centering
\includegraphics[width=0.48\textwidth]{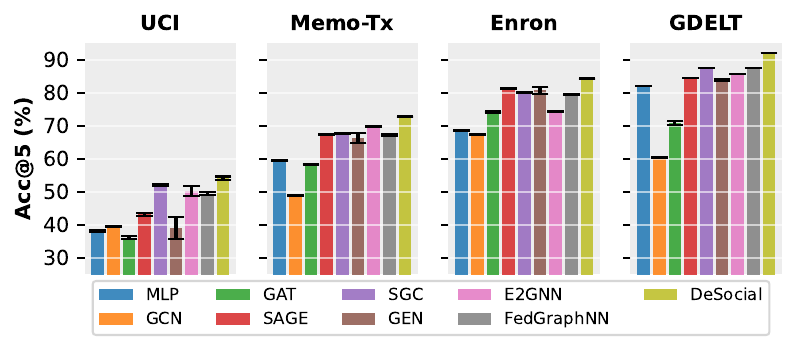}
\caption{Performance of \model and baselines measured in Acc@5 (\%). The mean and standard deviation across five validator settings runs are reported.}
\label{fig:acc5}
\end{figure} 

\begin{figure}[!t]
\centering
\includegraphics[width=0.48\textwidth]{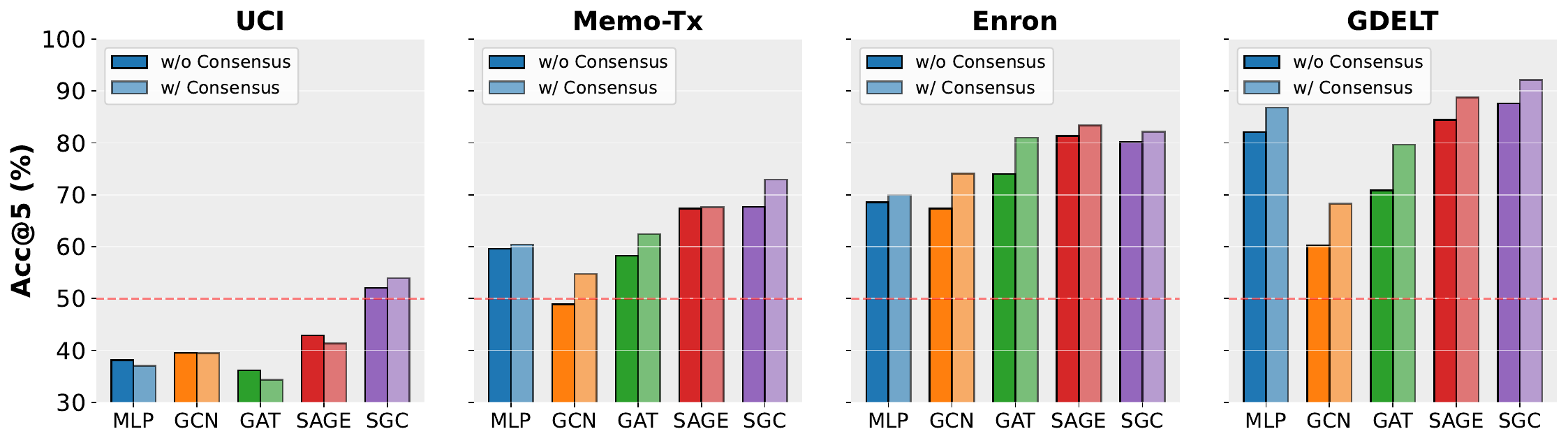}
\caption{Acc@5 of centralized vs. distributed recommendations, with 50\% dashed line indicating the random voting. Majority voting amplifies both strong and weak backbones.}
\label{fig:acc5_cons}
\end{figure}

\subsection{Implementation Details}

\begin{algorithm}[ht]
    \renewcommand{\baselinestretch}{0.1}\normalsize
    \raggedright
    \SetAlgoLined
    \LinesNotNumbered
    \caption{Rule-based Backbone Selection \label{alg:rule_selection}}
    \KwIn{Node degree $Deg(u)$, clustering coefficient $c_u$, backbone pool $\mathcal{F}$}
    \KwOut{Selected model $\mathcal{F}_u$ for node $u$}

    \uIf{$Deg(u)\ge6$ \&\& SGC$\in\mathcal{F}$}{\Return SGC}
    \uElseIf{$c_u<0.2$ \&\& $Deg(u)\ge4$ \&\& SAGE$\in\mathcal{F}$}{\Return SAGE}
    \uElseIf{$Deg(u)\le2$ \&\& MLP$\in\mathcal{F}$}{\Return MLP}
    \uElseIf{$c_u\ge0.4$ \&\& GCN$\in\mathcal{F}$}{\Return GCN}
    \Else{\Return last model in $\mathcal{F}$}
\end{algorithm}

\label{sec:imp_detail}

\textbf{Hyper-Parameters in Algorithm Selection}. We run the algorithm selection by setting the following hyperparameters. Time decay coefficient $\alpha \in \{0, -0.01, -0.1, -1\}$ explores increasing levels of decay. $\alpha=0$ means no decay. More negative values discount older interactions faster. Sampled pairs $\gamma$ denote the number of positive–negative neighbor connection pairs per user in backbone evaluation. Larger $\gamma$ provides more negative samples, improving reliability. In our experiments, we set $\gamma \in \{250, 500, 750, 1000, 1250\}$. The recommendation algorithm pool $\mathcal{F}\in 2^{\mathcal{F}_{Full}}$ is a subset of the full set $\mathcal{F}_{Full}$ (i.e., MLP, GCN, GAT, GraphSAGE, and SGC). 

\textbf{Hyper-Parameters Setting}. For \textbf{MLP}, \textbf{GCN}, \textbf{GAT}, \textbf{GraphSAGE} and \textbf{SGC}, we conduct an extensive hyperparameter search for a fair comparison. Concretely, we search the learning rate from $\{1e-1, 5e-2, 1e-2, 5e-3, 1e-3, 5e-4, 1e-4, 5e-5, 1e-5, 5e-6, 1e-6\}$ and dropout from $\{0.3, 0.5, 0.7\}$. For each backbone on each dataset, we pick the config with the best validation performance for the corresponding metric and report the performance on the test set accordingly. Each model is trained for 100 epochs with early stopping patience set to 20 epochs. For \textbf{GEN}, we tune the residual connection coefficient from $0$ to $1$ with a step size of $0.1$. For \textbf{E2GNN}, we adapt it to link prediction and train the teacher and student networks for 100 epochs. Both the reward for the agent and the loss for the student are defined purely in terms of the KL divergence with respect to the teacher distribution, once the associated weighting hyperparameters are tuned. For \textbf{FedGraphNN}, we applied the subgraph-level federated learning methods and divided the training set equally to five clients in each period. We set the local training epochs in $\{2,5\}$ and aggregate the model parameters using FedAvg~\citep{mcmahan2017communication}, training 100 epochs in total. To ensure that the results of different baselines are reproducible and comparable, \textbf{GEN}, \textbf{E2GNN} and \textbf{FedGraphNN} are set with the same hyperparameter range and use the best configuration for each backbone. For \textbf{GEN} and \textbf{E2GNN}, we report the result with the best combination of backbones ranged in {\text{MLP}, \text{GCN}, \text{GAT}, \text{GraphSAGE}, \text{SGC}}. For \textbf{FedGraphNN}, which supports a single backbone per run, we train five variants, each using one of {\text{MLP}, \text{GCN}, \text{GAT}, \text{GraphSAGE}, \text{SGC}} as the client backbone.

\textbf{PyG Implementation}.
For GCN, GAT, and GraphSAGE, we implemented the models by applying two \texttt{GCNConv}, \texttt{GATv2Conv} and \texttt{SAGEConv} in \texttt{PyG}\footnote{\href{https://pytorch-geometric.readthedocs.io/en/latest/index.html}{https://pytorch-geometric.readthedocs.io/en/latest/index.html}}
 respectively, and we used dot products for decoding. In GAT, we used 4 head attention instead of 8 to save computation memory and time, and added BatchNorm. For implementing SGC, we used \texttt{SGConv} for encoding and multiperception layers for decoding. The graph training algorithms are implemented based on the open-source DTGB benchmark~\cite{zhang2024dtgb}.

\end{document}